\documentclass{emulateapj}
\bibpunct{(}{)}{;}{a}{}{,}

\begin{document}
\submitted{Received 2007 February 28; accepted 2007 April 04}

\title{Hydrodynamical simulations of the jet in the symbiotic star MWC\,560 \\
III. Application to X-ray jets in symbiotic stars}

\author{Matthias Stute and Raghvendra Sahai}
\affil{Jet Propulsion Laboratory, California Institute of Technology, 
4800 Oak Grove Drive, Pasadena, CA 91109, USA}

\email{Matthias.Stute@jpl.nasa.gov}

\begin{abstract}
In papers I and II in this series, we presented hydrodynamical simulations 
of jet models with parameters representative of the symbiotic system MWC\,560. 
These were simulations of a pulsed, initially underdense jet in a high 
density ambient medium. Since the pulsed emission of the jet creates internal 
shocks and since the jet velocity is very high, the jet bow shock and the 
internal shocks are heated to high temperatures and should therefore emit 
X-ray radiation. In this paper, we investigate in detail the X-ray properties 
of the jets in our models. We have focused our study on the total X-ray 
luminosity and its temporal variability, the resulting spectra and the spatial 
distribution of the emission. Temperature and density maps from our 
hydrodynamical simulations with radiative cooling presented in the second 
paper are used together with emissivities calculated with the atomic database 
ATOMDB. The jets in our models show extended and variable X-ray emission 
which can be characterized as a sum of hot and warm components with 
temperatures that are consistent with observations of CH Cyg and R Aqr. 
The X-ray spectra of our model jets show emission line features which 
correspond to observed features in the spectra of CH Cyg. The innermost parts 
of our pulsed jets show iron line emission in the 6.4 -- 6.7 keV range which 
may explain such emission from the central source in R Aqr. We conclude that 
MWC\,560 should be detectable with Chandra or XMM-Newton, and such X-ray 
observations will provide crucial for understanding jets in symbiotic stars. 
\end{abstract}


\section{Introduction}

Highly collimated fast outflows or jets are common in many astrophysical 
objects of different sizes and masses: active galactic nuclei (AGN), X-ray 
binaries (XRBs), young stellar objects (YSO), pre-planetary nebulae (PPN), 
supersoft X-ray sources and symbiotic stars. In the last two objects, the jet 
engine consists of an accreting white dwarf. In symbiotic stars, the 
companion is a red giant undergoing strong mass loss. More than one hundred 
symbiotic stars are known, but only about ten systems show the presence of 
jets. The most famous systems are R Aquarii, CH Cygni, and MWC\,560.

R Aquarii, with a distance of about 200 pc, is one of the nearest symbiotic 
stars and a well known jet source. The jet has been extensively observed in 
the optical and at radio wavelengths \citep[e.g.][]{SoU85,PaH94,HMK85,HLD85}.
It shows a rich morphology, e.g. a series of parallel features in the jet and 
the counter-jet, extending to a few hundred AU each. This is a hint of pulsed 
ejection of both jets. Furthermore, R Aqr is the first jet in a symbiotic 
system, which was detected in X-rays \citep{VPF87,HSS98,KPL01}. \citet{KPL01} 
found peaks of \ion{O}{7} at 0.57 keV in both the NE and SW jets and a peak of 
\ion{N}{6} at 0.43 keV only in the NE jet. The spectra are consistent with a 
soft component with $k\,T \sim$ 0.25 keV. The central source shows a supersoft 
blackbody emission with $k\,T \sim$ 0.18 keV and a Fe K$\alpha$ line at 6.4 
keV which suggests the presence of a hard source near the hot star. Recently, 
\citet{KAD06} reported on five years of observations with Chandra and 
were able to measure the proper motion of knots in the NE jet of about 600 km 
s$^{-1}$. \citet{NDK07} investigated the X-ray emission from the inner 500 AU
of this system.

In 1984/85, CH Cygni showed a strong radio outburst, during
which a double-sided jet with multiple components was ejected \citep{TSM86}.
This event allowed an accurate measurement of the jet velocity near 1500 km 
s$^{-1}$. In HST observations \citep{EBS02}, arcs can be detected that also 
could be produced by episodic ejection events. X-ray emission was first 
detected by EXOSAT \citep{LeT87}, and subsequent ASCA observations 
revealed a complex X-ray spectrum with two soft components ($k\,T =$ 0.2 and 
0.7 keV) associated with the jet, an absorbed hard component (7.3 keV) and a 
Fe K$\alpha$ line \citep{EIM98}. They interpreted the hard component as thermal
emission by material being accreted on to the white dwarf and the soft 
component as either coronal emission from the giant star or emissions from 
shocks in the jets. Analysis of archival Chandra ACIS data by 
\citet{GaS04} revealed faint extended emission to the south, aligned with the 
optical and radio jets seen in HST and VLA observations. 
\citet{WhK06} reanalyzed the ASCA data and interpreted the soft emission as 
scattering of the hard component in a photo-ionized medium surrounding the 
white dwarf. They claim that no other sources than the accreting white dwarf 
are required to explain the spectrum. The obvious existence of the jet, 
however, and furthermore the apparent decline of the hard X-ray component 
observed with the US-Japanese X-ray satellite Suzaku by \citet{MIK06} together 
with the lack of a corresponding decline in the soft component, suggest that
this interpretation is implausible. Recently, \citet[][hereafter KCR07]{KCR07} 
reported the detection of multiple components, including an arc, in the 
archival Chandra images.

R Aqr and CH Cyg are the only two jets of symbiotic stars which are detected 
in X-rays. While these two objects are seen at high inclinations, in MWC\,560 
the jet axis is practically parallel to the line of sight. This special 
orientation allows one to observe the outflowing gas as line absorption in 
the source spectrum. With such observations the radial velocity and the column 
density of the outflowing jet gas close to the source has been investigated in 
great detail. In particular the acceleration and evolution of individual 
outflow components and jet pulses has been probed with spectroscopic 
monitoring programs, as described in \citet{SKC01}. Using this optical 
data, sophisticated numerical models of this pulsed propagating jet have been 
developed \citep[][hereafter Paper I and II in this series]{SCS05,Stu06}. A 
number of hydrodynamical simulations 
(with and without cooling) were made in which the jet
density and velocity during the pulses were varied. The basic model
absorption line profiles in MWC\,560 as well as the mean velocity and
velocity-width are in good agreement with the observations. The evolution of
the time-varying high velocity absorption line-components is also well modeled.
These models not only fit the MWC\,560 data, but are also able to explain
properties of jets in other symbiotic systems such as the observed velocity
and temperature of the CH Cyg jet.

So far, MWC\,560 has not been detected in X-rays \citep{MWJ97}. We find using
the PIMMS tool\footnote{http://heasarc.gsfc.nasa.gov/Tools/w3pimms.html}  that 
the non-detection in the ROSAT all-sky survey sets an upper limit of the 
absorbed X-ray flux of 0.07 counts s$^{-1}$ \citep{MWJ97} and 
$7\times10^{-13}$ erg s$^{-1}$ cm$^{-2}$, respectively. 

The jets in all three symbiotic stars show evidence of episodicity. Such 
episodicity in the ejection process has been seen in numerical models of 
the interaction of the stellar magnetosphere and the accretion disk 
\citep[e.g.][]{GWB97,MGW02}. In the disk-wind scenario 
\citep[e.g.][]{BlP82,ALK03} the time-dependent emission is created by time 
variations in the accretion rate of the underlying disk.
Unfortunately, so far no hydrodynamical models exist for explaining the X-ray 
emission from symbiotic stars. As a first step we have therefore used our 
existing simulations, which fit MWC\,560, for understanding the observed X-ray 
emission properties of MWC\,560, CH Cyg and R Aqr. 

In \S \ref{sec_model}, we briefly describe the numerical models we have used. 
The total X-ray luminosity and its time dependence is examined in \S 
\ref{sec_lum}. After the resulting spectra are calculated in \S 
\ref{sec_spec}, we show emission maps in \S \ref{sec_maps} and apply our 
main results to X-ray observations of CH Cyg, MWC\,560 and R Aqr in \S 
\ref{sec_obs}. Finally our conclusions are given in \S \ref{sec_concl}.

\section{The numerical models} \label{sec_model}

\subsection{The hydrodynamical simulations}

We solved the equations of ideal hydrodynamics with an additional cooling term 
in the energy equation 
\begin{eqnarray} \label{hydro}
\frac{\partial\,\rho}{\partial\,t} + \nabla\,(\rho\,{\bf v}) &=& 0 \nonumber \\
\frac{\partial\,(\rho\,{\bf v})}{\partial\,t} + \nabla\,
(\rho\,{\bf v}\otimes{\bf v}) &=& - \nabla\,p \nonumber - \rho\,\nabla\,\Phi\\
\frac{\partial\,e}{\partial\,t} + \nabla\,(e\,{\bf v}) &=& - p\,\nabla\,{\bf v}
 - n^2\,\Lambda ( T ) \nonumber \\
p &=& (\gamma - 1)\,e .
\end{eqnarray}
\noindent
with the code {\em NIRVANA\_CP} which was written by 
\citet{ZiY97} and modified by \citet{Thi00} to calculate radiative losses due 
to non-equilibrium cooling by line emission. $\rho$ is the gas density, $p$ 
the pressure, $e$ the internal energy density, $\Phi$ the gravitational 
potential, ${\bf v}$ the velocity and $\gamma$ the ratio of the specific heats 
at constant pressure and volume which is set to $\gamma = 5/3$. The general 
capabilities of the code have been described in detail in Paper I, for our 
approximations and assumptions related to the cooling treatment we refer the 
reader to Paper II.  We briefly describe the geometry which we have adopted in 
our simulations below. 

We use a cylindrical coordinate system where the jet axis corresponds to the
z axis and both components of the binary system are located in the plane 
perpendicular to this axis. The hot component is located at the origin of the 
coordinate frame; with a binary separation of 4 AU, a red giant is implemented.
The red giant is surrounded by a stellar wind with constant velocity of 10 km 
s$^{-1}$ and a mass loss rate of 10$^{-6}$ M$_{\odot}$ yr $^{-1}$.

The jet is produced within a thin jet nozzle with a radius of 1 AU. The 
initial velocity of the steady jet is chosen to 1000 km s$^{-1}$ 
and its density is set to $8.4 \times 10^{-18}$ g cm$^{-3}$ (equal to a 
hydrogen number density of $5 \times 10^6$ cm$^{-3}$). These parameters
lead to i) a mass loss rate of $\approx 10^{-8}$ M$_{\odot}$ yr $^{-1}$ in the 
steady jet, and ii) a density contrast between the steady jet and the ambient 
medium $\eta$ of $5 \times 10^{-3}$ and a Mach number of $\approx 60$ in the 
jet nozzle at the origin of the coordinate system. Repeatedly each seventh 
day, the velocity and density values in the nozzle are changed to simulate the 
jet pulses which are seen in the observations of MWC\,560. The duration of each
pulse is one day.

Two models (iv' and i') out of our existing set of eight models were chosen 
for computing X-ray emission properties. These models represent maximum  
(model iv') and minimum (model i') values of the jet density in the pulses. 
In model iv' (model i'), the jet density in the pulses is higher (lower) than 
the jet density in the steady jet. Although model iv' provided the best fit 
for the optical data for MWC\,560, our work in this paper shows that model i' 
results in X-ray properties which are more appropriate for CH Cyg. For both 
models we used an approximate treatment of radiative cooling. The model 
properties including the velocities and densities of the jet pulses are given 
in Table \ref{pulses}. 

\begin{deluxetable*}{lcccccc}
\tablecaption{Parameters of the jet pulses \label{pulses}}
\tablewidth{\textwidth}
\tablehead{
\colhead{model} & 
\colhead{$n_{\rm{pulse}}$ [cm$^{-3}$]} & 
\colhead{$v_{\rm{pulse}}$ [km s$^{-1}$]} &
\colhead{$\dot M_{\rm jp}$ [$\mbox{M}_{\odot}$ yr$^{-1}$]} &
\colhead{$L_{\rm jet}$ [erg s$^{-1}$]}}
\startdata
i' & $1.25 \times 10^6$ & $2000$ & 
$4.66 \times 10^{-9}$ & $5.88 \times 10^{33}$ \\
iv' & $1 \times 10^7$ & $2000$ & 
$3.73 \times 10^{-8}$ & $4.70 \times 10^{34}$ \\
\enddata
\tablecomments{The values of the steady jet emission are $n = 5 \times 10^6$ 
cm$^{-3}$, $v = 1000$ km s$^{-1}$, $\dot M_{\rm js} = 9.33 \times 10^{-9}$ 
$\mbox{M}_{\odot}$ yr$^{-1}$ and $L_{\rm jet} = 2.93 \times 10^{33}$ erg 
s$^{-1}$, respectively. $\dot M_{\rm js}$ and $\dot M_{\rm jp}$ are the mass 
outflow rates of the jet in the steady state and during the pulse, 
respectively. The duration of each pulse is one day, their period is seven 
days.}
\end{deluxetable*}

\begin{figure*}
\plotone{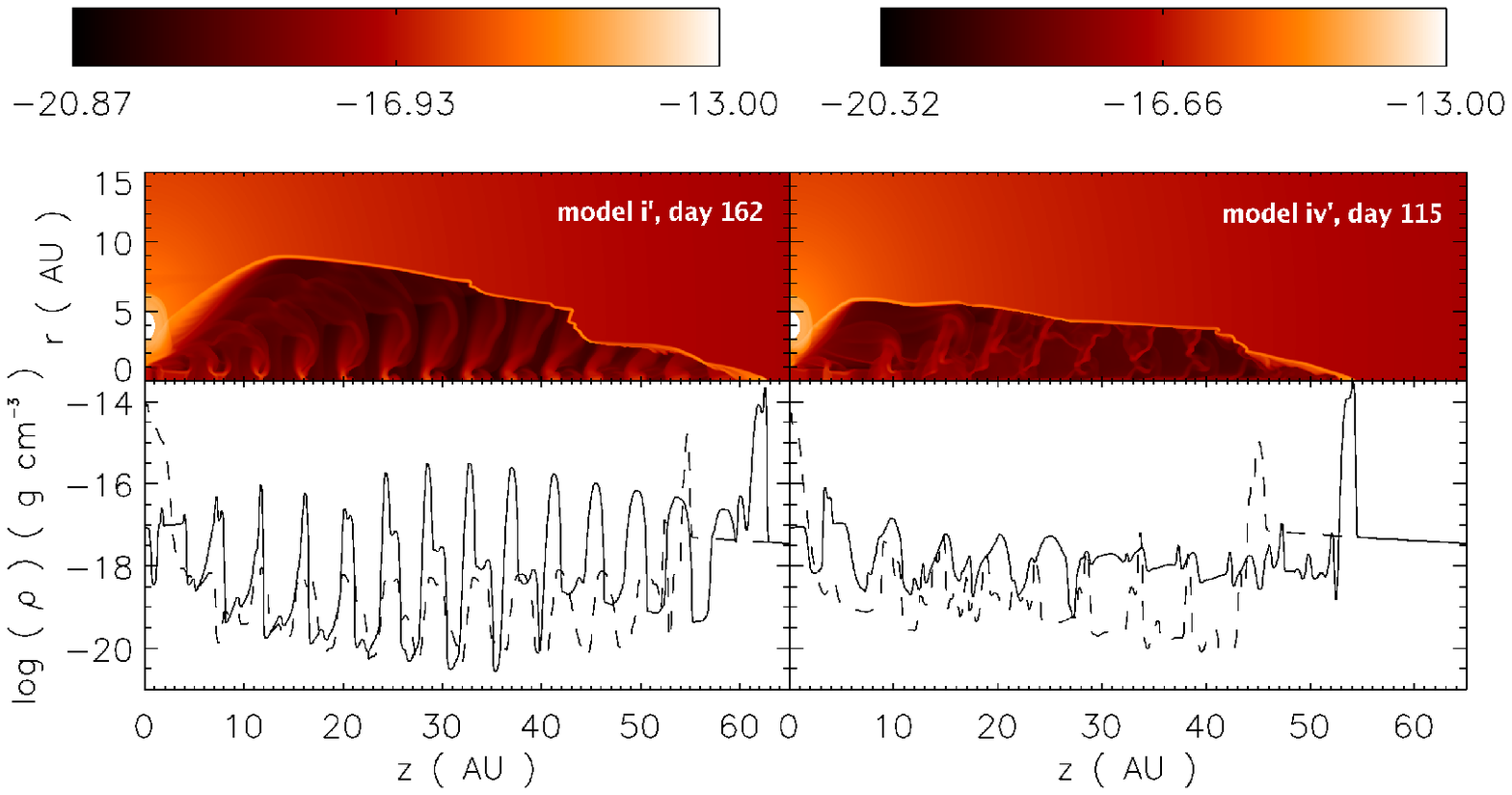}
\plotone{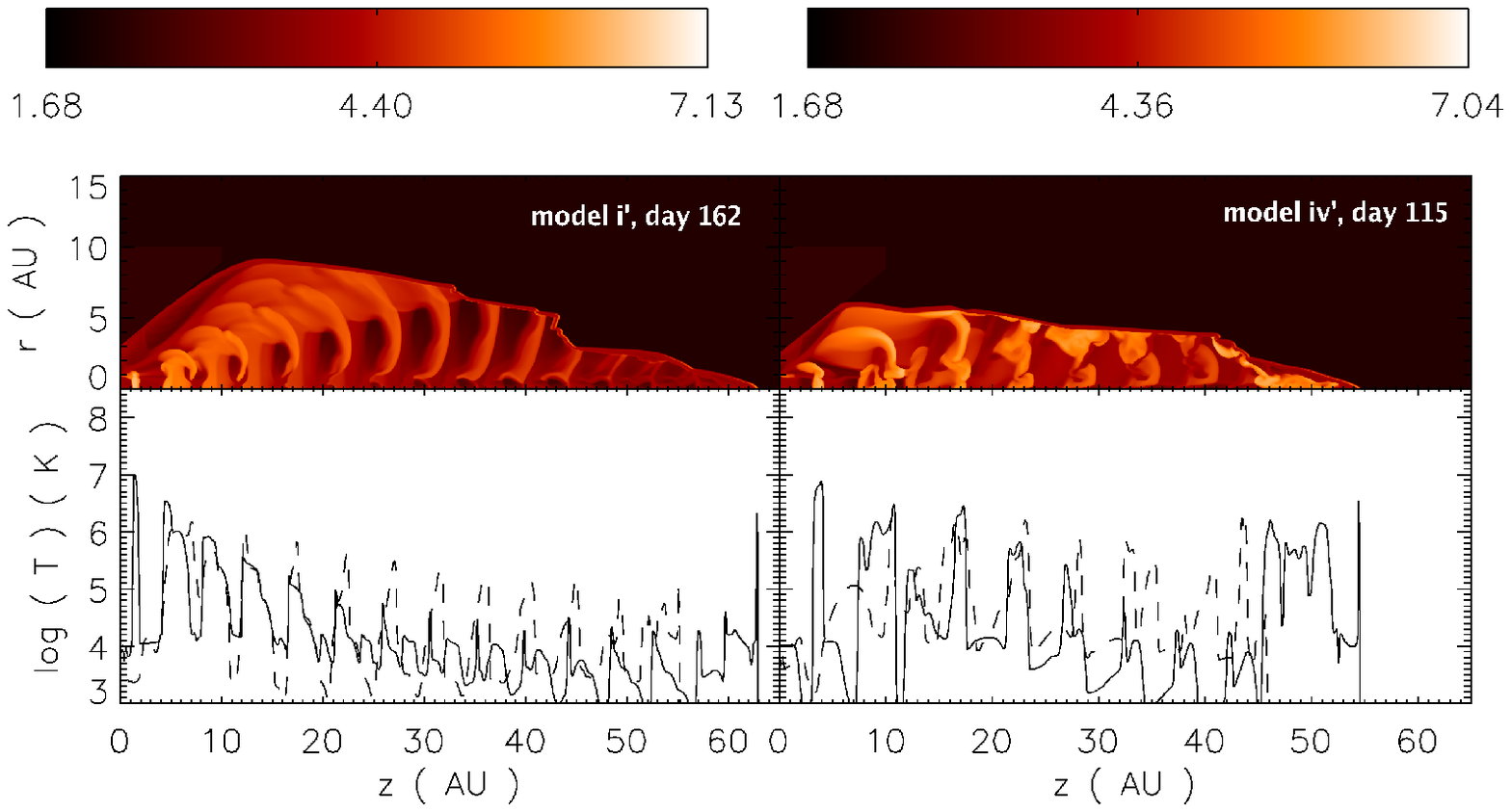}
\caption{Logarithm of density (top) and temperature (bottom) for model i' 
on day 162 (left) and for model iv' on day 115 (right), shown as contour plots
and slices along the jet axis (solid) and parallel to the jet axis at $r = 2$ 
AU (dashed).}
\label{den_temp}
\end{figure*}

Maps of logarithm of density and temperature for model i' on day 162 and for 
model iv' on day 115 are given in Fig. \ref{den_temp}. Both are the last 
time-steps calculated.

\subsection{Calculating the X-ray properties}

We determined the expected X-ray flux using the density and temperature maps 
from the hydrodynamical simulations as follows. We used
the atomic database ATOMDB with IDL including the Astrophysical
Plasma Emission Database (APED) and the spectral models output from the
Astrophysical Plasma Emission Code \citep[APEC,][]{SBL01} to calculate 
the emissivity. The default abundances in ATOMDB, i.e. 14 elements (H, He, C,
N, O, Ne, Mg, Al, Si, S, Ar, Ca, Fe, Ni) with solar abundances of 
\citet{AnG89}, are used. The energy range is divided into bins of 0.01 keV.  
We compute the spectrum and the total flux in X-rays as a function of 
evolutionary time for each of our models.

We calculate the X-ray emission in the range between 0.15 -- 15 keV,
which is exactly the energy range covered by EPIC on XMM-Newton
and includes that of the ACIS instrument (0.2 -- 10 keV) and of HETG (0.4 -- 
10 keV) on Chandra. The emission from gas with a temperature lower than 
$\lesssim 10^6$ K is only marginal in this energy range. 

\section{The total X-ray luminosity and its time dependence} \label{sec_lum}

\begin{figure}
\plotone{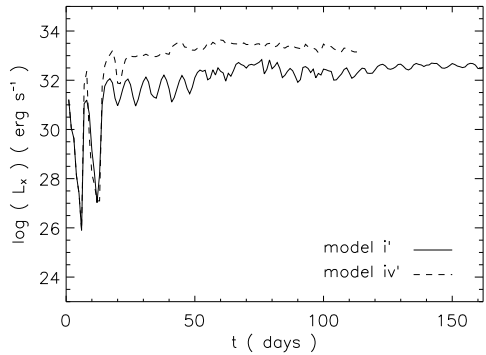}
\caption{X-ray luminosity as a function of the evolution time of the jet 
in model i' (solid) and model iv' (dashed) in the energy range 0.15 -- 15 keV. 
Most of the energy is emitted below 2 keV, i.e. the plots showing the 
luminosity in the energy range 0.15 -- 2 keV are similar.}
\label{Xray_lum}
\end{figure}

As expected, the high temperatures, created by the interaction of the jet 
pulses with previously ejected matter, lead to substantial X-ray emission
(Fig. \ref{Xray_lum}). The X-ray luminosity in model iv' is higher than in 
model i'. This is a result of the higher density in the pulses and thus of the 
higher kinetic luminosity $L_{\rm jet} = 1/2 \, \dot M \, v^2$ pumped into the 
jet. Furthermore, in model i' about 5\% of the average kinetic luminosity is 
radiated in X-rays, but in model iv' about 19\%. Since the X-rays are emitted
by shocked material from the fast moving pulses and since the X-ray luminosity
is proportional to $\rho^2$, compared to $L_{\rm jet}$ being proportional 
to $\rho$, the ratio of the X-ray luminosity to the kinetic luminosity is 
proportional to $\rho$. Therefore this ratio is higher in model iv' than in 
model i'.

\begin{figure}
\plotone{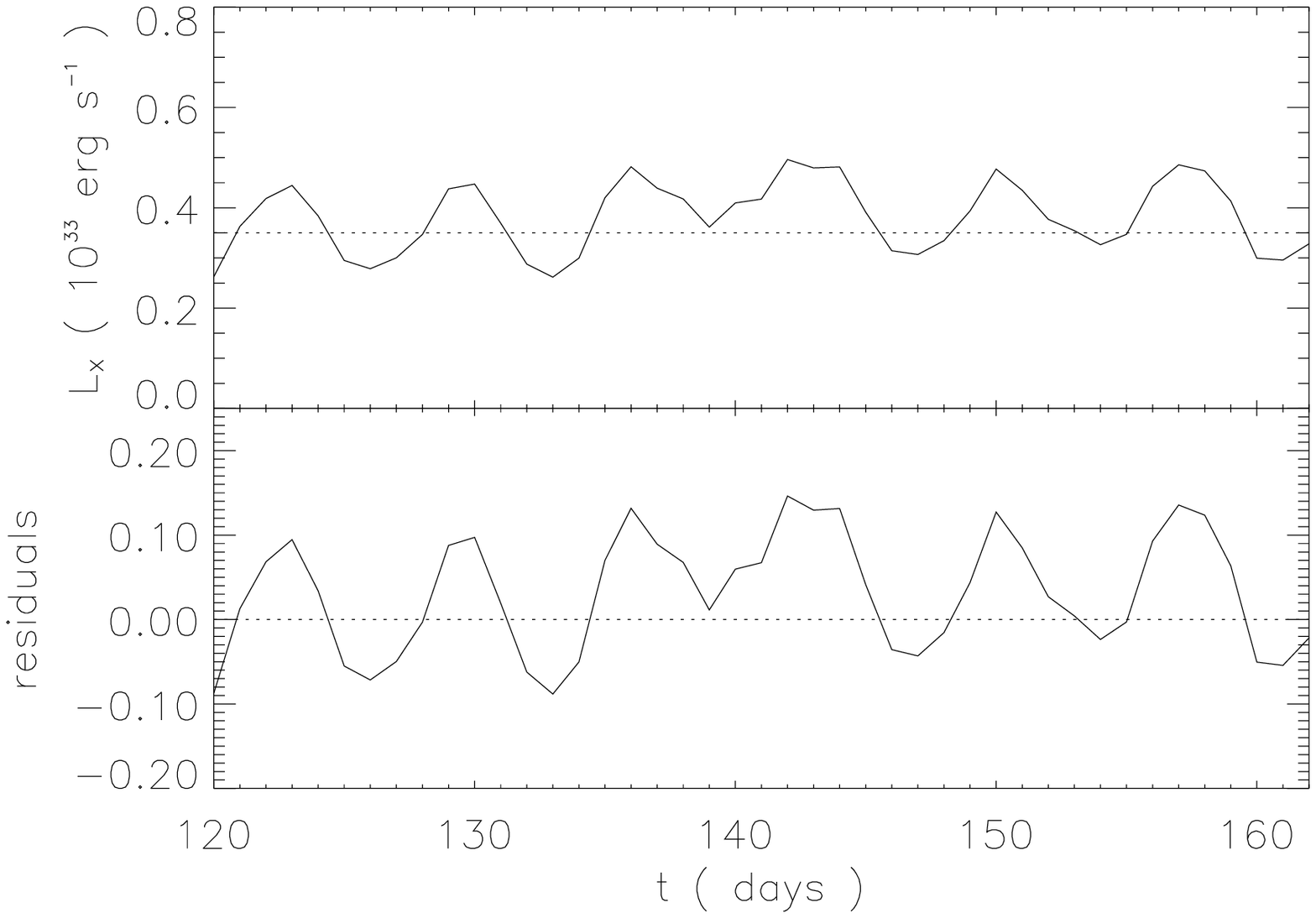}
\plotone{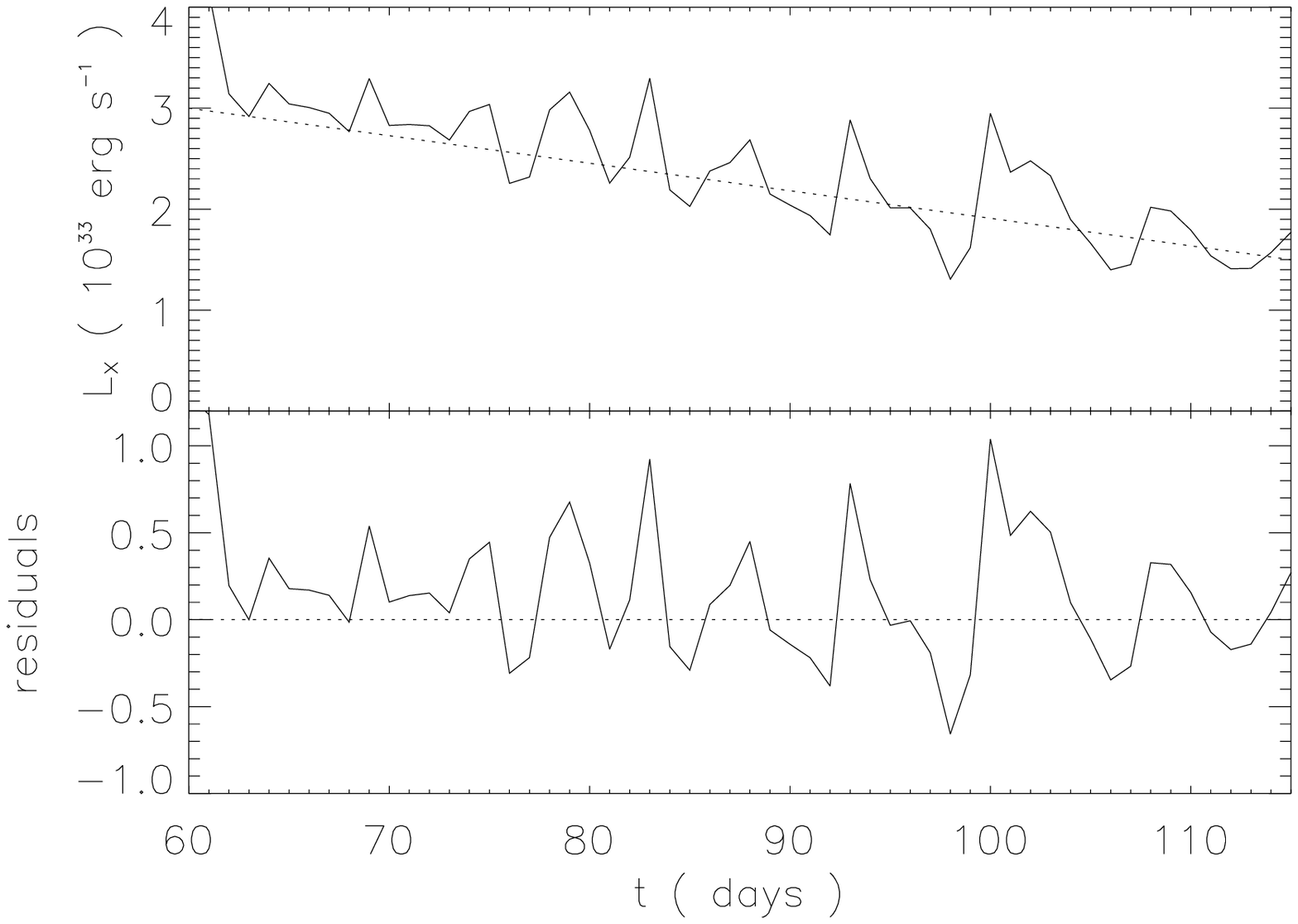}
\caption{X-ray luminosity as a function of time  
for model i' (top) and model iv' (bottom); one can see 
minima and maxima created by the emergence of each new jet pulse; the dotted
lines in the top plots show the different trends of the X-ray 
luminosity decreases with time.}
\label{flashes}
\end{figure}

We find minima and maxima in the X-ray emission $L_{\rm X}$ (computed by 
integrating over the energy range 0.15 -- 15 keV) which are connected with the 
periodic emergence of jet pulses (Fig. \ref{flashes}). Thus
the period of the variations in the X-ray emission is about 7 days. The size of
the fluctuations is 50 \% and more of the average emission. 

While the X-ray luminosity stays constant with time for model i', it decreases 
with time for model iv'. This difference might be related to a larger amount 
of cooling in model iv'. The initial shock temperature is identical in both 
models, since the velocities are the same. The higher densities in the jet 
pulses in model iv', however, lead to higher densities in the X-ray emitting 
material and thus to higher pressures which result in stronger adiabatic 
expansion and hence enhanced adiabatic cooling. Radiative cooling is also 
enhanced by the higher densities in model iv'.

\section{The spectrum and its time dependence} \label{sec_spec}

The spectra of both models in the energy range between 0.15 -- 15 
keV show many different features. They show continuum emission, and 
superimposed on the continuum, a large number of emission features (some of 
which are blends of several emission lines). A prominent feature, which is 
mainly due to blended iron lines, is seen between 0.7--1 keV. 
Iron also produces a strong emission feature in the 
6.4 -- 6.7 keV range requiring very high temperatures ($\sim$ $10^7$ K) 
that are reached locally in the jet. 

\begin{figure}
\plotone{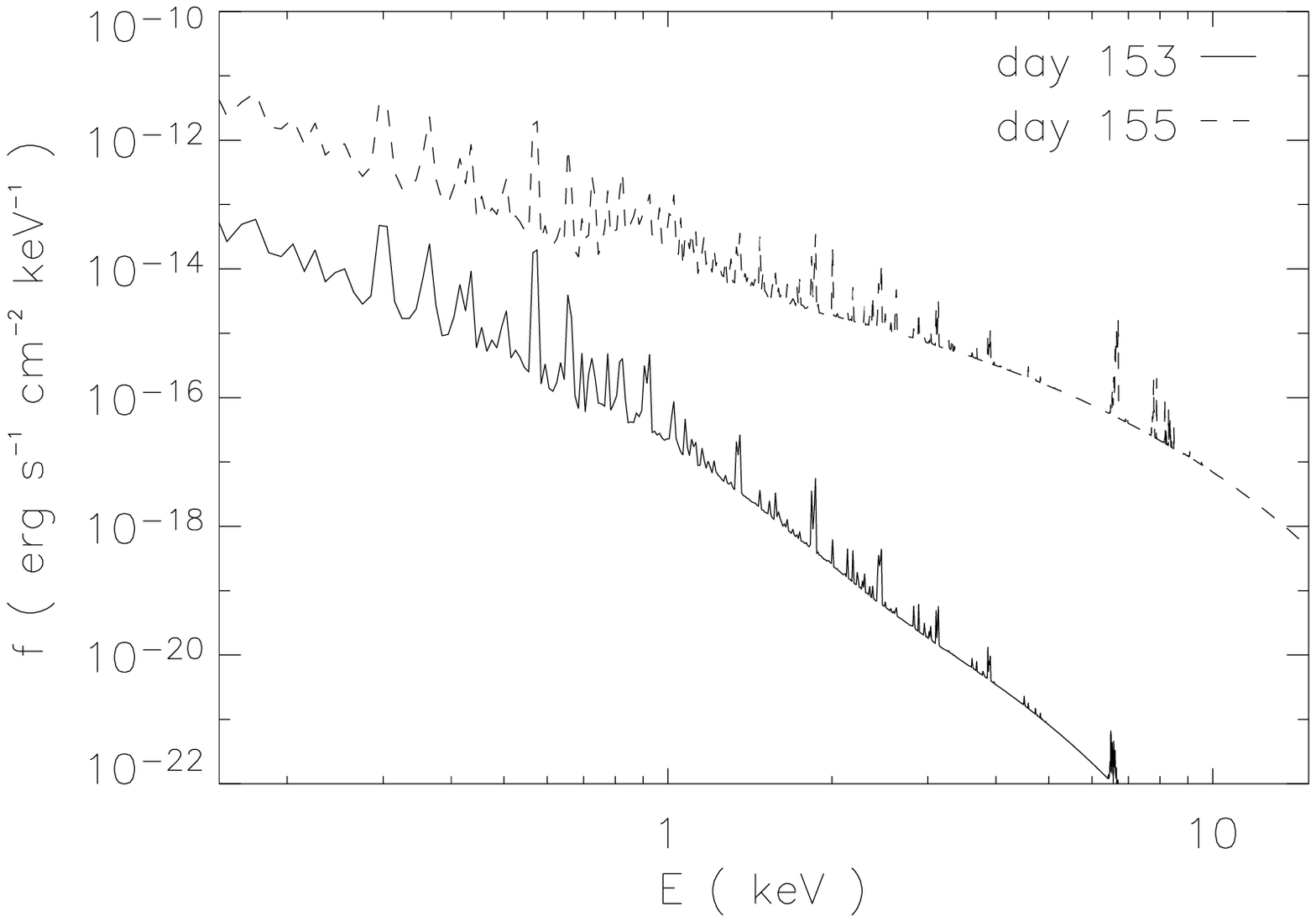}
\plotone{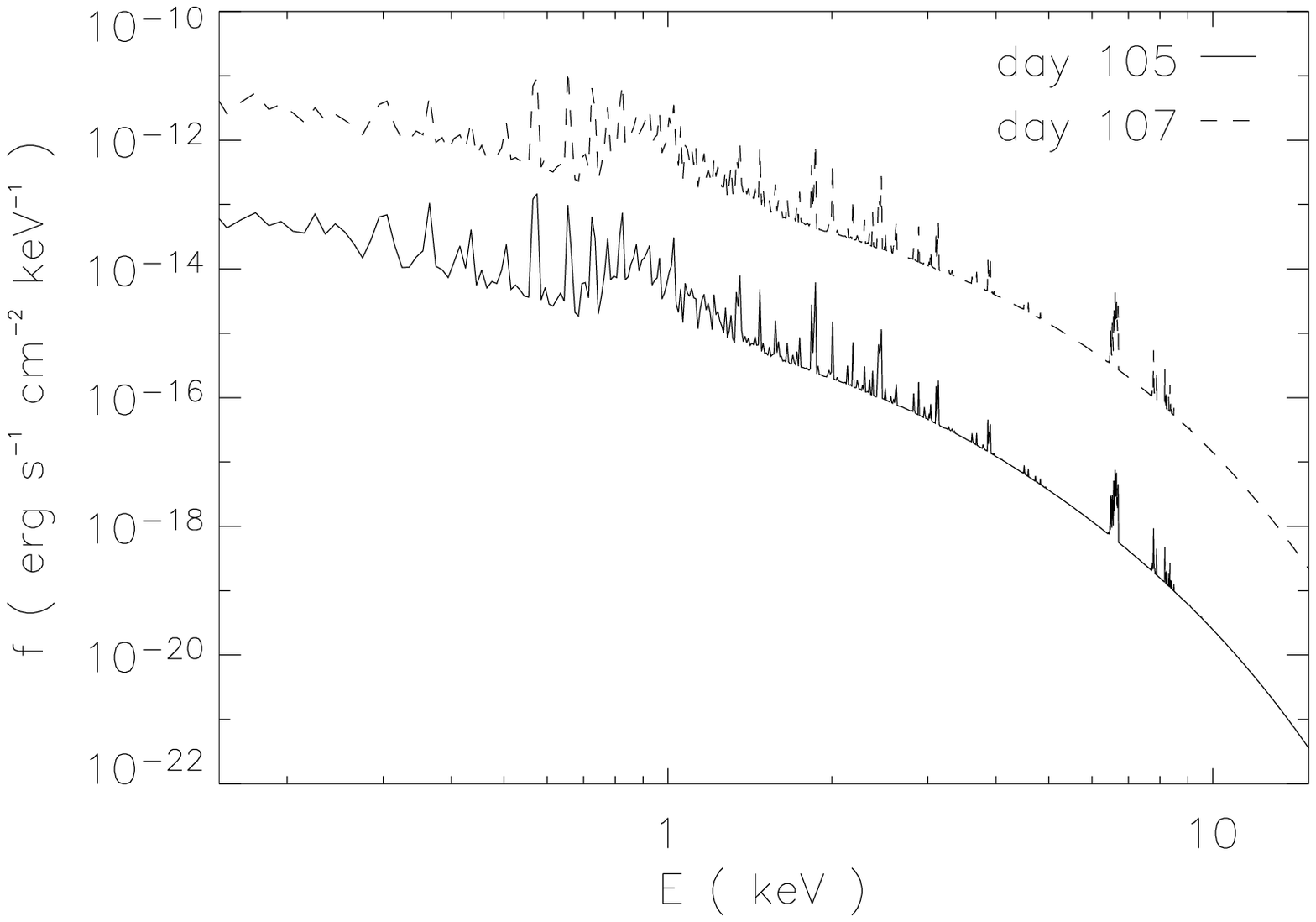}
\caption{Spectrum in the energy range between 0.15 -- 15 keV; top: for 
model i' on days 153 (a minimum of the total X-ray luminosity, solid) and 155 
(close to the maximum X-ray luminosity, dashed), bottom: for model iv' on 
days 105 (solid) and 107 (dashed). The spectrum plotted with a solid line is 
shifted downwards by a factor of 100 for clarity in each plot. One can clearly 
see the time dependence of the spectrum.}
\label{fullspectrum}
\end{figure}

Like the total X-ray luminosity, the spectrum is also highly time-dependent 
(Fig. \ref{fullspectrum}). We define two proxies for the temperature, one using
the low energy spectrum and one using the high energy spectrum. These proxies
can then be used conveniently for direct comparison with the single-temperature
thermal plasma models typically used to fit the observed data. 

\subsection{Defining temperature proxies}

\begin{figure}[htb]
\plotone{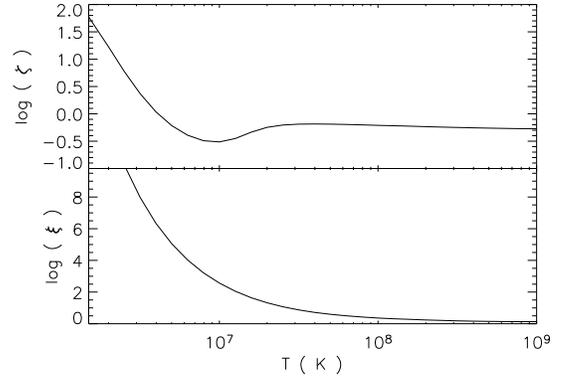}
\caption{Temperature proxies as a function of temperature for a 
single-temperature thermal plasma calculated with ATOMDB: hardness ratio 
$\zeta$ (top) and flux ratio $\xi$ (bottom) as defined in the text.}
\label{proxiesvsT}
\end{figure}

In order to characterize the temperature of the propagating jet from the 
low energy spectrum, we use the fact that below energies of about 0.7 keV, both
spectra in Fig. \ref{fullspectrum} are almost identical, but differ 
significantly between 0.7 and 2 keV. Therefore we define the proxy $\zeta$ for 
relatively low temperature plasma ($10^7$ K) in the jet as
\begin{equation}
\zeta = \frac{f_{\rm (0.2-0.7)\,keV}}{f_{\rm (0.7-2)\,keV}}
\end{equation}
with $f_{\rm (0.2-0.7)\,keV}$ and $f_{\rm (0.7-2)\,keV}$ the fluxes integrated 
over the given energy ranges (Fig. \ref{proxiesvsT}, top). 

The slope of the continuum also changes with time. To measure this behavior, we
define a flux ratio 
\begin{equation}
\xi = \frac{f_{\rm 4\,keV}}{f_{\rm 9\,keV}}
\end{equation}
We choose regions in the spectrum where no lines are present, although photons
with these high energies have not been observed from the jet or might be 
confused with photons from the central engine in observed spectra.

\subsection{Determining the temperature from model spectra} 
\label{sec_temp_mod}

\begin{figure}
\plotone{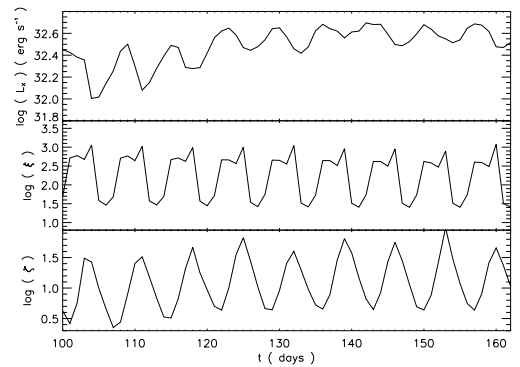}
\caption{X-ray luminosity of the jet as a function of time in 
the energy range 0.15 -- 15 keV (top), flux ratio $\xi$ as a 
function of time (middle) and hardness ratio $\zeta$ as a 
function of time (bottom) for models i'.}
\label{fluxratio1}
\end{figure}

In model i', the high temperature proxy, $\log (\xi)$, varies with time 
spanning a range from 1.5 to 3. It has its highest value (i.e. the spectrum has
the steepest slope, hence the average temperature is at its lowest value), when
the X-ray luminosity shows a minimum (Fig. \ref{fluxratio1}). Comparison of 
the model $\log (\xi)$ values with that of a single-temperature thermal
plasma (Fig. \ref{proxiesvsT}, bottom) gives us temperature estimates of the 
hot component between 
$8\times10^6$ K (0.69 keV) and $1.7\times10^7$ K (1.5 keV). The highest 
temperatures are consistent with that of post-shock gas with a shock velocity 
of about 1100 km s$^{-1}$. The minimum in the $\log (\xi)$ (i.e. maximum in 
average temperature) coincides with the emergence of each new pulse; within 
the next 2--3 day period the compressed knot cools and the emissivity 
increases. Therefore the maximum in the X-ray luminosity is reached about 2--3 
days later. 

The low temperature proxy, $\log (\zeta)$, varies between about 0.5 and 1.6, 
the corresponding temperatures lie between $1.6\times10^6$ K (0.14 keV) and 
$3\times10^6$ K (0.26 keV). The jet is therefore better described as a 
combination of a warm and a hot component rather than as a single-temperature 
plasma. 

\begin{figure}
\plotone{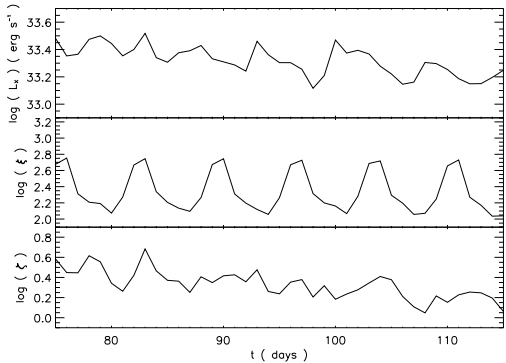}
\caption{same as Fig. \ref{fluxratio1}, but for model iv'}
\label{fluxratio2}
\end{figure}

In model iv', $\log (\xi)$ varies between 2.1 and 2.8; the corresponding
temperatures are $8\times10^6$ K (0.69 keV) and $1.2\times10^7$ K (1.03 keV). 
The low temperature proxy, $\log (\zeta)$, lies between 0.1 and 0.6; the
corresponding temperatures are $3\times10^6$ K (0.26 keV) and $3.8\times10^6$ K
(0.33 keV). As in model i', the jet is better characterized as a combination of
a warm and a hot component. 

The range of temperatures in the hot component over one pulse cycle in model 
iv' is smaller compared to that in model i'; this is because the higher density
in model iv' makes radiative cooling more efficient, such that the shock 
heating is damped more efficiently. The different density contrasts between the
jet pulses and the steady jet in both models also lead to different 
shock velocities and thus to different shock temperatures to which the plasma 
is heated initially.

\subsection{Emission lines in the model spectra}

\begin{figure}
\plotone{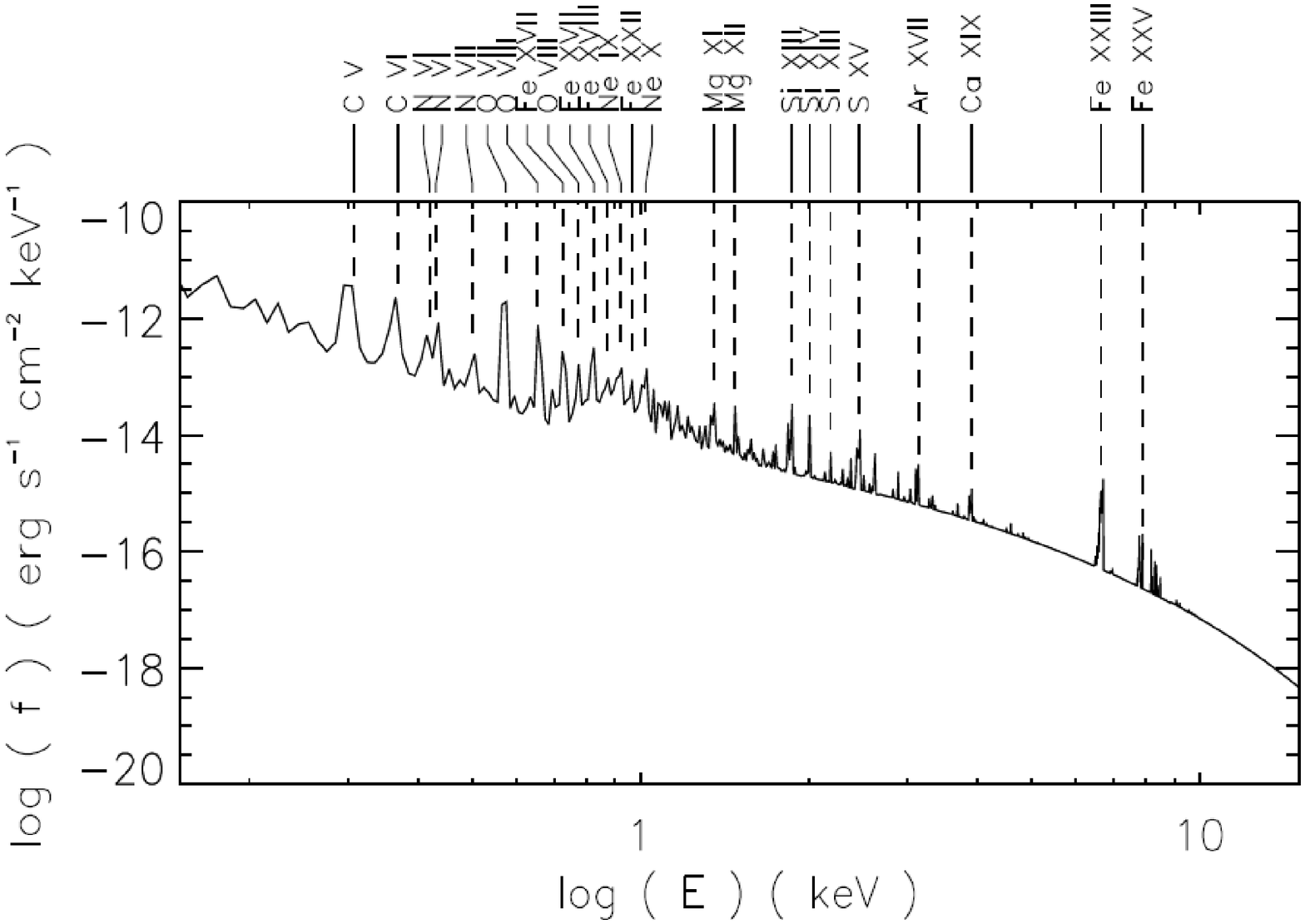}
\caption{Spectrum in the energy range between 0.15 -- 15 keV for 
model i' on days 155; several strong emission lines are present with very 
different ionization potentials which can only be explained using two 
components at different temperatures.}
\label{lines}
\end{figure}

We have identified the strongest emission lines characterizing a typical jet 
spectrum using the spectrum derived from model i' on day 155 as a template 
(Fig. \ref{lines}). The most prominent lines are the oxygen line at 0.57 keV, 
the Ne lines at 0.93 and 1.03 keV, the Mg lines at 1.35 and 1.47 keV, the Si 
lines at 1.86, 2.01 and 2.18 keV, and the Fe complex at about 6.5 keV. All 
strong lines with their identifications are given in Table \ref{linestable}, 
including their fluxes at days 153 and 155 in model i' and days 105 and 107 in 
model iv'. Most of these lines are hydrogenic or He-like lines of heavy 
elements, however, also lines of higher ionization states are present 
(\ion{Fe}{22}, \ion{Fe}{23}, \ion{Fe}{25}). Thus this set of lines with very 
different ionization potentials can also only explained with two components at 
different temperatures.  

\begin{table*}
\centering
\caption{Prominent emission lines in the model spectra\label{linestable}}
\tablewidth{\textwidth}
\begin{tabular}{cccccc}
\hline
\hline
ion & energy (keV) & \multicolumn{4}{c}{flux (erg s$^{-1}$ cm$^{-2}$)} \\ 
& & \multicolumn{2}{c}{model i'} & \multicolumn{2}{c}{model iv'} \\
& & day 153 & day 155 & day 105 & day 107 \\
\hline
\ion{C}{5}   & 0.31 & 8.81$\times10^{-14}$ & 7.02$\times10^{-14}$ 
& 1.36$\times10^{-13}$ & 7.03$\times10^{-14}$ \\
\ion{C}{6}   & 0.37 & 3.09$\times10^{-14}$ & 2.49$\times10^{-14}$ 
& 1.06$\times10^{-13}$ & 4.21$\times10^{-14}$ \\
\ion{N}{6}   & 0.42 & 6.96$\times10^{-15}$ & 6.53$\times10^{-15}$ 
& 2.58$\times10^{-14}$ & 1.08$\times10^{-14}$ \\
\ion{N}{6}   & 0.43 & 1.00$\times10^{-14}$ & 9.20$\times10^{-15}$ 
& 4.06$\times10^{-14}$ & 1.50$\times10^{-14}$ \\
\ion{N}{7}   & 0.50 & 2.59$\times10^{-15}$ & 3.07$\times10^{-15}$ 
& 2.31$\times10^{-14}$ & 1.68$\times10^{-14}$ \\
\ion{O}{7}   & 0.57 & 3.80$\times10^{-14}$ & 3.67$\times10^{-14}$ 
& 2.67$\times10^{-13}$ & 1.59$\times10^{-13}$ \\
\ion{O}{8}   & 0.65 & 5.57$\times10^{-15}$ & 9.26$\times10^{-15}$ 
& 1.11$\times10^{-13}$ & 1.37$\times10^{-13}$ \\
\ion{Fe}{17} & 0.73 & 5.32$\times10^{-16}$ & 3.70$\times10^{-15}$ 
& 9.17$\times10^{-14}$ & 8.93$\times10^{-14}$ \\
\ion{Fe}{17} & 0.83 & 5.73$\times10^{-16}$ & 4.11$\times10^{-15}$ 
& 9.34$\times10^{-14}$ & 9.39$\times10^{-14}$ \\
\ion{O}{8}   & 0.77 & 4.19$\times10^{-16}$ & 1.65$\times10^{-15}$ 
& 2.89$\times10^{-14}$ & 3.13$\times10^{-14}$ \\
\ion{Fe}{18} & 0.87 & 2.17$\times10^{-17}$ & 7.42$\times10^{-16}$ 
& 2.29$\times10^{-14}$ & 2.68$\times10^{-14}$ \\
\ion{Ne}{9}  & 0.92 & 5.90$\times10^{-16}$ & 1.99$\times10^{-15}$ 
& 3.60$\times10^{-14}$ & 5.00$\times10^{-14}$ \\
\ion{Fe}{22} & 0.97 & 1.30$\times10^{-17}$ & 7.52$\times10^{-16}$ 
& 1.41$\times10^{-14}$ & 2.58$\times10^{-14}$ \\
\ion{Ne}{10} & 1.02 & 8.72$\times10^{-17}$ & 1.70$\times10^{-15}$ 
& 3.76$\times10^{-14}$ & 4.17$\times10^{-14}$ \\
\ion{Mg}{11} & 1.35 & 3.34$\times10^{-17}$ & 4.00$\times10^{-16}$ 
& 9.41$\times10^{-15}$ & 1.07$\times10^{-14}$ \\
\ion{Mg}{12} & 1.47 & 1.71$\times10^{-18}$ & 2.72$\times10^{-16}$ 
& 4.28$\times10^{-15}$ & 7.17$\times10^{-15}$ \\
\ion{Si}{13} & 1.86 & 5.25$\times10^{-18}$ & 3.86$\times10^{-16}$ 
& 7.04$\times10^{-15}$ & 1.06$\times10^{-14}$ \\
\ion{Si}{14} & 2.01 & 3.71$\times10^{-19}$ & 2.12$\times10^{-16}$ 
& 1.37$\times10^{-15}$ & 4.30$\times10^{-15}$ \\
\ion{Si}{13} & 2.18 & 4.51$\times10^{-19}$ & 3.81$\times10^{-17}$ 
& 5.76$\times10^{-16}$ & 9.99$\times10^{-16}$ \\
\ion{S}{15}  & 2.46 & 1.35$\times10^{-18}$ & 1.43$\times10^{-16}$ 
& 1.55$\times10^{-15}$ & 3.34$\times10^{-15}$ \\
\ion{Ar}{17} & 3.14 & 6.06$\times10^{-20}$ & 3.31$\times10^{-17}$ 
& 2.03$\times10^{-16}$ & 5.60$\times10^{-16}$ \\
\ion{Ca}{19} & 3.9  & 1.52$\times10^{-20}$ & 1.05$\times10^{-17}$ 
& 3.68$\times10^{-17}$ & 1.30$\times10^{-16}$ \\
\ion{Fe}{23} & 6.65 & 2.11$\times10^{-20}$ & 4.01$\times10^{-17}$ 
& 4.81$\times10^{-18}$ & 2.01$\times10^{-17}$ \\
\ion{Fe}{25} & 7.78 & 1.17$\times10^{-23}$ & 2.76$\times10^{-18}$ 
& 1.17$\times10^{-18}$ & 7.12$\times10^{-18}$ \\
\hline
\end{tabular}
\end{table*}

\section{The X-ray emission maps} \label{sec_maps}

\begin{figure*}
\plotone{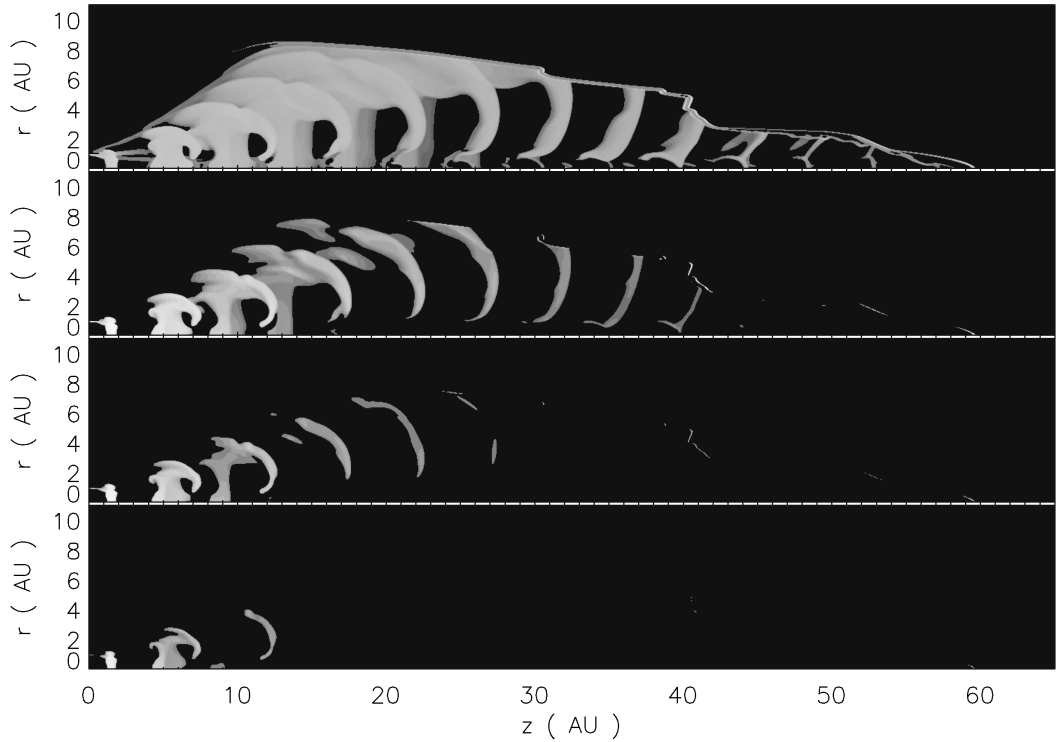}
\caption{Emission maps of model i' on day 155 in the 0.15 -- 1.5 keV range,
in the 1.5 -- 3 keV range, in the 3 -- 6 keV range and in the 6 -- 7 keV 
range.}
\label{maps}
\end{figure*}

The results in the previous section suggest the existence of two components,
a warm one with temperatures in the range of (1.6 -- 3.8)$\times 10^6$ K and 
a hot one with temperatures of (8 -- 17)$\times 10^6$ K, respectively. As 
already pointed out in Paper I and II, the jet consists of dense, cool knots 
and tenuous, hot inter-knot gas. The knots are too cold to emit X-rays, thus 
the low and high energy components in the X-ray spectrum probe the temperature 
structure of the hot parts of the jet (Fig.1 and Fig. 12 in paper I). These 
hot parts consist of shocked material in the inter-knot gas segments. 
Temperature gradients are present within each inter-knot gas segment and also 
between the inter-knot gas close to the jet source and those downstream near 
the jet head. The regions with the highest temperatures and thus emission 
above 6 keV lie in the first two inter-knot gas segments within only a few AU 
from the central source (Fig. \ref{maps}). The inter-knot gas segments become 
progressively cooler, as they move farther away from the jet source due to 
adiabatic expansion and the resulting cooling of jet material. 

\section{Comparison with observations} \label{sec_obs}

We now compare the results of our models to X-ray observations of the three 
objects MWC\,560 (only upper limits to the flux are available), CH Cyg and R 
Aqr.

\subsection{MWC\,560}

The source fluxes in the 0.2 -- 2.4 keV range are $3 \times 10^{-13}$ erg 
s$^{-1}$ cm$^{-2}$ for model i' and $2 \times 10^{-12}$ erg s$^{-1}$ cm$^{-2}$ 
for model iv', using a distance of 2.5 kpc to MWC\,560
\citep[][and references therein]{SKC01}. The latter flux is reduced to an 
absorbed flux of $1.7 \times 10^{-13}$ erg s$^{-1}$ cm$^{-2}$, using the model 
of the visual interstellar extinction in the Galaxy by \citet{HJS97} which 
gives an $A_{\rm v} = 0.88$ or $N_{\rm H} = 1.55 \times 10^{21}$ cm$^{-2}$ for 
MWC\,560. The absorbed flux is consistent with MWC\,560's non-detection 
in the ROSAT all-sky survey \citep[$<7 \times 10^{-13}$ erg 
s$^{-1}$ cm$^{-2}$,][]{MWJ97}. The fluxes, however, are high enough such that 
today's X-ray telescope as Chandra and XMM-Newton should be able to detect 
MWC\,560\footnote{XMM-Newton observations proposed by the authors will be 
executed in AO-6 after May 2007.}.

\subsection{CH Cyg}

Since the jet velocities are similar in MWC\,560 and CH Cyg, we can make a 
detailed comparison of our models with the latter object. This comparison is 
limited to the soft emission from the propagating jet and excludes the X-ray 
emission above about 2 keV which is believed to be dominated by the variable 
scattered hard X-rays from the central source. 

\citet{EIM98} resolved for the first time atomic emission lines (or their 
blends) from elements in hydrogenic and He-like ionization states in the 
X-ray spectra of CH Cyg. The most prominent features in the observed spectra 
are the oxygen line at 0.57 keV, the blend of i) Ne lines at 0.93 and 1.03 keV,
ii) Mg lines at 1.35 and 1.47 keV, iii) Si lines at 1.86, 2.01 and 2.18 keV, 
and the Fe line complex at about 6.5 keV. These lines are also seen in our 
model spectra. Furthermore, they also had to introduce a two-temperature 
thermal plasma model to explain the set of lines detected in the ASCA spectrum.
The temperatures of their two components (0.21 and 0.72 keV) are in the same 
range as in our model i' (warm component: 0.14 -- 0.26 keV, hot component: 
0.69 -- 1.5 keV, see \S \ref{sec_temp_mod}). In model iv', the warm component 
is too hot (0.26 -- 0.33 keV) to explain the observations.

Recently, KCR07 reported the detection of multiple components, 
including an arc, in the archival Chandra images. This arc has a similar 
opening angle to many of the arcs visible in our X-ray emission maps (Fig. 
\ref{maps}), and its presence supports our pulsed-jet model in which such 
emission results from internal shocks generated by colliding jet ejecta.

Considering the smaller distance of 268 pc to CH Cyg, the model X-ray 
luminosity of $2.2\times10^{32}$ erg s$^{-1}$ for model i' and 
$1.5\times10^{33}$ erg s$^{-1}$ for model iv' in the 0.2 -- 2 keV range 
implies fluxes of $2.6 \times 10^{-11}$ erg s$^{-1}$ cm$^{-2}$ and $1.74 
\times 10^{-10}$ erg s$^{-1}$ cm$^{-2}$, respectively. 
The interstellar extinction for CH Cyg is $A_{\rm v} = 0.006$ or 
$N_{\rm H} = 1 \times 10^{20}$ cm$^{-2}$, thus the absorbed flux is $2 \times 
10^{-11}$ erg s$^{-1}$ cm$^{-2}$ for model i' and $1.3 \times 10^{-10}$ erg 
s$^{-1}$ cm$^{-2}$ for model iv', respectively. The measured fluxes of the 
soft components associated with the jet, however, lie in the range $(0.4-3.8)
\times 10^{-12}$ erg s$^{-1}$ cm$^{-2}$ \citep{EIM98,GaS04,MIK06}. Since these
measured fluxes are smaller than those predicted by model i' and by far
smaller than those from model iv' which has a higher jet pulse density than 
model i', we infer that the jet pulse densities in CH Cyg are smaller than 
those in MWC\,560 and our models. If, as stated in \S \ref{sec_lum}, the 
ratio of the X-ray luminosity to the kinetic energy pumped into the jet is 
proportional to the jet pulse density, we can estimate that the jet pulse 
density has to be reduced to about $10^5$ cm$^{-3}$ to model the X-ray flux 
observed in CH Cyg. In order to resolve the difference between measured and 
model fluxes with uncertainties in the distance to CH Cyg, the latter would 
have to be about 700 pc, however, this large value is very unlikely, since the 
distance was measured with HIPPARCOS with an error of 23 \%.

KCR07 give an observational estimate of the density in the jet, 50 
cm$^{-3}$, based on the total X-ray luminosity, assuming an emitting sphere 
with a radius of 400 AU and a mean emissivity of $2\times10^{-23}$ erg cm$^3$ 
s$^{-1}$. However, astrophysical jets, typically, are collimated structures 
which have significantly smaller volumes than a sphere. Since we simulated the 
propagation of the model jets only up to a length of 65 AU, not 400 AU, we 
scale up its volume as follows. If we assume that the cross-sectional 
area stays constant as the jet propagates and evolves with time, the jet volume
is about ($\pi\,\times6^2\times400$) AU$^{3}$ (Fig. \ref{maps}) which is 
smaller by about a factor of 6000 than that of a sphere with a radius of 400 
AU. Alternately if we assume that the lateral expansion of the jet is 
proportional to its axial expansion\footnote{which is seen in our models after 
day 70 (Paper II)}, the volume is still overestimated by a factor of about 160.
Furthermore, within the volume of the jet, the X-rays are emitted by clumps and
not by the whole jet, i.e. the filling factor for the X-ray emission is, 
$f_1<1$. In addition, KCR07 assume a temperature of $1.2\times 10^7$ K 
(1 keV) to estimate the emissivity, however, our jet models show that such high
temperatures are only archieved in the innermost region of the jet; at 
larger distances from the jet source along the axis, the X-ray emitting knots 
are significantly cooler (Fig. \ref{den_temp}). Thus the value of the 
emissitivity should be lower and the density should be higher than estimated 
by KCR07 by another correction factor $f_2>1$. Hence we conclude that an 
accurate estimate of the X-ray emitting volume and the temperature would lead 
to a higher density by a factor of (160--6000)$f_2/f_1$ compared to the value 
given by KCR07, of the order of $10^4$--$10^6$ cm$^{-3}$. This range
compares well with the densities of material emitting soft X-rays in our 
models which are of the order of $10^5$--$10^6$ cm$^{-3}$ (Fig. 
\ref{den_temp}).

Other possibilities to reduce the X-ray fluxes in our models, bringing
them closer to the observed ones in CH Cyg, are i) a longer timescale between 
the pulses and ii) a smaller velocity difference between the steady jet and 
the jet pulses. In the first case, less energy is pumped into the jet and each 
jet knot and the jet head can cool further before being hit by the next pulse. 
In the second case, the smaller velocity difference reduces the temperature 
to which the shocked material is heated. However, the density contrast then 
also has to be adjusted in order to reproduce the observed proper motion of 
the jet knots. Which of the above scenarios is the most likely one can only 
be determined from future simulations which have been fine-tuned to fit the 
properties of the X-ray emitting material in the jet of CH Cyg. 

In 1994 and 2006, the flux from the jet in CH Cyg was almost at the same level,
in the range $(2.7-3.8)\times 10^{-12}$ erg s$^{-1}$ cm$^{-2}$ 
\citep{EIM98,MIK06}. In 2001, however, it was lower by a factor of ten 
\citep{GaS04}. In the context of our models, such a drop in flux may result 
from a large decrease in the density of the pulses which may be caused by a 
drop in the accretion rate onto the white dwarf. 

\subsection{R Aqr}

Since the jet velocity in R Aqr is smaller by a factor of 2 than those in our 
models, we can only make a more limited comparison of the model results with 
the observations. \citet{KAD06} and also \citet{KKS06} report a tangential 
motion of an X-ray emitting knot\footnote{\citet{KKS06} give an even smaller 
velocity of 380 km s$^{-1}$ in their Table 1, however, without explaining the 
difference from the value given in the text.} of 600 km s$^{-1}$. 
\citet{KKS06} estimated a density of 100 cm$^{-3}$, however, it may be possible
that, as in the case of CH Cyg, the X-ray emitting volume is overestimated and
thus the density is underestimated.

\citet{KPL01} found in R Aqr that the NE jet is more luminous by a 
factor of 3 than the SW jet. The spectrum the NE jet was fitted with a 
single-temperature thermal plasma with a temperature of 1.66 keV, the spectrum 
of the SW jet with a plasma temperature of 0.2 keV. The simplest explanation of
this difference is that the ambient media on both sides have different 
densities which would lead to different compression factors and thus to 
different shock heating temperatures. However, proper motion measurements show 
no significant differences in the velocities of the knots in both sides of the 
jet \citep{PaH94, MLV04}, which we would expect if different ambient densities 
would decelerate the jet differently on both sides.

Our modeling can provide a plausible explanation for the observed differences 
between the two sides of the jet, if we assume that the ejection of the jet 
pulses on both sides are out of phase with each other. \citet{HOM91} 
derived a kinematic age of both jets of about 90 yrs; over the extent of the 
jet we can observe 3--6 knots in the observations of e.g. \citet{PaH94} which 
suggests several ejection events in this period of time. A period of about 17 
years for these ejection events has been inferred from radio observations 
\citep{KHY89}, thus they are significantly larger than in MWC\,560 and in our 
models\footnote{The events in R Aqr are thought to be triggered by periastron 
passage, while the variations in MWC\,560 seem to be a result of disk 
instabilities.}. We hypothesize that the X-ray emitting blobs in the NE jet 
were ejected later than those in the SW jet and therefore have cooled less. 
This hypothesis is supported by the fact that the X-ray emitting component in 
the NE jet is closer to the central core (about 8'') than the blobs in the SW 
jet \citep[12--26'',][]{KPL01}. A new SW jet component with an offset of about 
1.5'' from the central source has recently been reported by \citet{NDK07}. 
Assuming a jet velocity of about 600 km s$^{-1}$ \citep{KAD06}, i.e. about 
0.6'' per year, we obtain a kinematic age of about 2.5 years for this 
component. Even if the new component had been ejected during the epoch of
\citet{KPL01}'s observations, its flux would have contributed to the central 
source, but not to that of the SW jet. Hence the presence of the new component 
does not conflict with our hypothesis. The time period between the emergence of
the new SW jet component and the emergence of the blob now located at 12'' is 
about 17 years which supports the inferred period for the ejection of jet 
pulses in R Aqr. 

In \S \ref{sec_lum}, we showed that the total X-ray luminosity decreases with 
time, probably due to adiabatic cooling in the jet (Paper II). This 
effect provides a plausible explanation of the decrease in X-ray flux in the 
jet of R Aqr from a value of $5 \times 10^{-13}$ erg s$^{-1}$ cm$^{-2}$ in the 
early 1990s \citep{HSS98} to $1\times 10^{-13}$ erg s$^{-1}$ cm$^{-2}$ in 2000 
\citep{KPL01}. If this explanation holds, we expect to see a further decrease 
in flux in future observations. 

\subsection{The 6.4 -- 6.7 keV iron line complex}

This iron line complex has been observed in both objects, CH Cyg and R Aqr.
Our model spectra also show the existence of this Fe line complex (\S 
\ref{sec_spec}). \citet{EIM98} fitted the observed spectrum of CH Cyg with 
three single-temperature components (a warm and a hot component to explain the 
jet emission and hard component for the central engine) and an additional 
Gaussian representing fluorescence emission in the Fe K$\alpha$ line. This 
fluorescence occurs close to the white dwarf and the accretion disk. Since it 
is not possible to disentangle the flux of the thermal and the fluorescence 
components in this line complex and since our models do not include the effect 
of fluorescence, we cannot compare our models with this part of the observed 
spectrum of CH Cyg.

In R Aqr, the origin of the hard X-ray emission is more ambiguous. 
\citet{KPL01} detected 16 photons in the range between 6.4 and 6.7 keV which 
they attribute to the central source due to the extraction regions they chose.
The physical origin of this emission, i.e. thermal or fluorescence, cannot be 
decided, since there are not enough photons to model the spectrum in this 
energy and characterize its nature. In our model, we find significant hard 
radiation including continuum and iron emission lines, being emitted by the 
first two internal shocks in the jet downstream from the source (i.e. at a 
distance less than 15 AU). Since at R Aqr's distance of 200 pc, 15 AU 
correspond to 75 milliarcseconds, which is well below the angular resolution 
of Chandra, the X-ray emission from these shocks cannot be separated from the
central source. The model X-ray luminosity in the 6.4--6.7 keV range is 
between $5.1 \times 10^{24}$ erg s$^{-1}$ and $7.3 \times 10^{28}$ erg 
s$^{-1}$ for model i' and between $2.8 \times 10^{28}$ erg s$^{-1}$ and 
$5.1 \times 10^{29}$ erg s$^{-1}$ for model iv', respectively, depending on 
whether the jet is in its minimum or maximum state. At a distance of 200 pc, 
this corresponds to fluxes between $1.1 \times 10^{-18}$ erg s$^{-1}$ cm$^{-2}$
and $1.1 \times 10^{-13}$ erg s$^{-1}$ cm$^{-2}$. Since \citet{KPL01} measured 
a flux of $4.9 \times 10^{-14}$ erg s$^{-1}$ cm$^{-2}$ at 6.41 keV, we 
suggest that the measured iron line flux may be emitted by the jet itself, and 
an additional fluorescence component is not needed, in R Aqr. However, new 
simulations with the same jet velocity as in R Aqr are needed in order to test 
this suggestion.

\section{Conclusions} \label{sec_concl}

We have used our models of pulsed, radiative jets in symbiotic stars in order 
to investigate their X-ray properties in detail. These models show that the 
well-studied pole-on jet in MWC\,560 should be easily detected by today's 
X-ray telescopes such as Chandra and XMM-Newton, since our model flux and its
time variation for this source are of the order of $10^{-13}$ erg s$^{-1}$ 
cm$^{-2}$. 

We find minima and maxima in the X-ray emission $L_{\rm X}$ (computed by 
integrating over the energy range 0.15 -- 15 keV) which are connected with the 
periodic emergence of jet pulses. The maxima of the total X-ray luminosity 
occur 2--3 days after the emergence of new jet pulses, which are ejected every 
7 days. The size of the fluctuations is 50 \% and more of the average 
emission, making such X-ray flashing jets detectable with Chandra and 
XMM-Newton.

The X-ray spectra of our model jets are rich in emission line features, the 
most prominent of which correspond to observed features in the spectra of CH 
Cyg. 

By using low and high energy temperature proxies derived from the spectra, we 
can show that the emission can be adequately characterized with a hot and a 
warm optically-thin plasma component. The hot component has temperature 
values of about 0.7 keV (1.6 keV) during the minima (maxima) of the total 
X-ray luminosity and the warm component has temperature values of about 0.14 
keV (0.33 keV) during the minima (maxima). 

While model iv' is appropriate for MWC\,560, we have shown that model i', which
has a lower jet pulse density than model iv', is more appropriate for the 
jet in CH Cyg in terms of explaining the lower X-ray flux. Other possibilities
to reduce the flux are a longer timescale between the pulses and a smaller 
velocity difference between the steady jet and the jet pulses. Which of the 
above scenarios is the most likely one has to be tested in future simulations 
which have to be tuned to the jet in CH Cyg. 

Our models also offer a plausible explanation for the differences in 
luminosities and temperatures in the NE and the SW jet of R Aqr. We assume 
that the ejection of the jet pulses on both sides are out of phase with each 
other. We hypothesize that the X-ray emitting blobs in the NE jet 
were ejected later than those in the SW jet and therefore have cooled less. 

We find the existence of iron line emission in the 6.4 -- 6.7 keV range in our 
models which is also observed in both, CH Cyg and R Aqr. Our models cannot be
directly applied to CH Cyg, because of an additional fluorescence component 
from the central source and accretion disk in the observed spectrum and because
fluorescence is not included in our models. In the case of R Aqr, although 
this emission has been associated with the central source, our modeling shows 
that it is consistent with being produced by jet knots very close to the 
latter, because their separation in our model is well below the angular 
resolution of Chandra.

Using our current models which were built to fit the optical data
of the jet in MWC\,560 we are able to explain some of the important
characteristics of X-ray emission from jets in MWC\,560 and other symbiotic 
stars. The results of this study demonstrate the great potential of future 
numerical simulations of pulsed jets which have been fine-tuned to specific
source properties for understanding the jet phenomenon in symbiotic stars.
Furthermore, we hope that this study will inspire new and more sensitive 
observations of X-ray emission from jets in symbiotic stars.

\acknowledgements
We acknowledge helpful and improving comments and suggestions by the referee. 
The hydrodynamical simulations were performed at the High Performance
Computing Center Stuttgart, Germany.  This work was partially funded by
NASA/CHANDRA grants GO3-4019X and GO4-5163Z, and NASA/STScI grant
HST-GO-10317.01-A. The research described in this publication was carried out 
at the Jet Propulsion Laboratory, California Institute of Technology, under a 
contract with the National Aeronautics and Space Administration.


\begin{thebibliography}{}
\bibitem[Anders \& Grevesse(1989)]{AnG89}
Anders, E., Grevesse, N. 1989, GeCoA 53, 197
\bibitem[Anderson et al.(2003)]{ALK03}
Anderson, J. M., Li, Z.-Y., Krasnopolsky, R., Blandford, R. D. 2003, \apj\ 590,
L107
\bibitem[Blandford \& Payne(1982)]{BlP82}
Blandford, R. D., Payne, D. G. 1982, \mnras\ 199, 883
\bibitem[Ezuka et al.(1998)]{EIM98}
Ezuka, H., Ishida, M., Makino, F. 1998, \apj\ 499, 388
\bibitem[Eyres et al.(2002)]{EBS02}
Eyres, S. P. S., Bode, M. F., Skopal, A., et al. 2002, \mnras\ 335, 526l
\bibitem[Galloway \& Sokoloski(2004)]{GaS04}
Galloway, D. K., Sokoloski, J. L. 2004, \apj\ 613, L61
\bibitem[Goodson et al.(1997)]{GWB97}
Goodson, A. P., Winglee, R. M., B\"ohm, K.-H. 1997, \apj\ 489,199
\bibitem[Hakkila et al.(1997)]{HJS97}
Hakkila, J., Myers, J. M., Stidham, B. J., Hartmann, D. H. 1997, \aj\ 114, 2043
\bibitem[Hollis et al.(1985a)]{HMK85}
Hollis, J.M., Michalitsianos, A. G., Kafatos, M., et al. 1985, \apj\ 289, 765
\bibitem[Hollis et al.(1985b)]{HLD85}
Hollis, J. M., Lyon, R. G., Dorband, J. E., et al. 1985, \apj\ 475, 231
\bibitem[Hollis et al.(1991)]{HOM91}
Hollis, J. M., Oliversen, R. J., Michalitsianos, A. G., et al. 1991, \apj\ 377,
227
\bibitem[Hollis et al.(1997)]{HPL97}
Hollis, J. M., Pedelty, J. A., Lyon, R. G. 1997, \apj\ 482, L85
\bibitem[H\"unsch et al.(1998)]{HSS98}
H\"unsch, M., Schmitt, J. H., Schroeder, K., Zickgraf, F. 1998, \aap\ 330, 225
\bibitem[Kafatos et al.(1989)]{KHY89}
Kafatos, M., Hollis, J. M., Yusef-Zadeh, F., et al. 1989, \apj\ 346, 991
\bibitem[Karovska et al.(2007)]{KCR07}
Karovska, M., Carilli, C. L., Raymond, J. C., Mattei, J. A.
2007, ApJ accepted, astro-ph/0703278
\bibitem[Kellogg et al.(2001)]{KPL01}
Kellogg, E., Pedelty, J. A., Lyon, R. G. 2001, \apj\ 563, 151
\bibitem[Kellogg et al.(2006)]{KAD06}
Kellogg, E., Anderson, C., DePasquale, J., et al. 2006, poster abstract 
\#70.08, American Astronomical Society Meeting 207
\bibitem[Korreck et al.(2006)]{KKS06}
Korreck, K. E., Kellogg, E., Sokoloski, J. L. 2006, in: "The Multicoloured 
Landscape of Compact Objects and their Explosive Origins", astro-ph/0611401
\bibitem[Leahy \& Taylor(1987)]{LeT87}
Leahy, D. A., Taylor, A. R. 1987, \aap\ 176, 262
\bibitem[M\"akinen et al.(2004)]{MLV04}
M\"akinen, K., Lehto, H. J., Vainio, R., Johnson, D. R. H. 2004, \aap\ 424, 157
\bibitem[Matt et al.(2002)]{MGW02}
Matt, S., Goodson, A. P., Winglee, R. M., B\"ohm, K.-H. 2002, \apj\ 574, 232
\bibitem[Mukai et al.(2006)]{MIK06}
Mukai, K., Ishida, M., Kilbourne, C., et al. 2006, \pasj\ accepted, 
astro-ph/0609245
\bibitem[M\"urset et al.(1997)]{MWJ97}
M\"urset, U., Wolff, B., Jordan, S. 1997 , \aap\ 319, 201
\bibitem[Nichols et al.(2007)]{NDK07}
Nichols, J. S., DePasquale, J., Kellogg, E., et al. 2007, \apj\ accepted
\bibitem[Paresce \& Hack(1994)]{PaH94}
Paresce, F., Hack, W. 1994, \aap\ 287, 154
\bibitem[Schmid et al.(2001)]{SKC01}
Schmid, H. M., Kaufer, A., Camenzind, M., et al. 2001, \aap\ 377, 206
\bibitem[Smith et al.(2001)]{SBL01}
Smith, R. K., Brickhouse, N. S., Liedahl, D. A., Raymond, J. C. 2001, \apj\
556, L91
\bibitem[Solf \& Ulrich(1985)]{SoU85}
Solf, J., Ulrich, H. 1985, \aap\ 148, 274
\bibitem[Stute(2006)]{Stu06}
Stute, M., 2006, \aap\ 450, 645 (Paper II)
\bibitem[Stute, Camenzind \& Schmid(2005)]{SCS05}
Stute, M., Camenzind, M., Schmid, H.~M. 2005, \aap\ 429, 209 (Paper I)
\bibitem[Sutherland \& Dopita(1993)]{SuD93}
Sutherland, R.~S., Dopita, M.~A. 1993, ApJS, 88, 253
\bibitem[Taylor et al.(1986)]{TSM86}
Taylor, A. R., Seaquist, E. R., Mattei, J. A. 1986, Nature, 319, 38
\bibitem[Thiele(2000)]{Thi00}
Thiele, M. 2000, Numerical simulations of protostellar jets, PhD Thesis,
University of Heidelberg
\bibitem[Viotti et al.(1987)]{VPF87}
Viotti, R., Piro, L., Friedjung, M., Cassatella, A. 1987, \apj\ 319, L7 
\bibitem[Wheatley \& Kallman(2006)]{WhK06}
Wheatley, P. J., Kallman, T. R. 2006, \mnras\ 372, 1602
\bibitem[Ziegler \& Yorke(1997)]{ZiY97}
Ziegler, U.,  Yorke, H. 1997, Comp. Phys. Comm. 101, 54
\end{thebibliography}
\end{document}